\documentstyle[preprint,aps,eqsecnum]{revtex}
\begin{document}
{
\tighten
\title {Difficulties in using the sharp neutrino spectrum at short times} 
\author{Harry J. Lipkin\,\thanks{Supported in part by 
U.S.
Department of Energy, Office of Nuclear Physics, under contract 
number
DE-AC02-06CH11357.}} 
\address{ \vbox{\vskip 0.truecm}
  Department of Particle Physics
  Weizmann Institute of Science, Rehovot 76100, Israel \\
\vbox{\vskip 0.truecm}
School of Physics and Astronomy,
Raymond and Beverly Sackler Faculty of Exact Sciences,
Tel Aviv University, Tel Aviv, Israel  \\
\vbox{\vskip 0.truecm}
Physics Division, Argonne National Laboratory,
Argonne, IL 60439-4815, USA\\
~\\harry.lipkin@weizmann.ac.il
\\~\\
}

\maketitle

\begin{abstract} 
Final states produced by a decay have a much broader energy spectrum than
the natural line width at times much shorter than the decay lifetime. This 
tends to render impossible the use for neutrino detection of the high value of
the resonance absorption cross section at the peak of the resonance.

\end{abstract}}  % end tighten 

Proposals to use the very sharp neutrino emission line for neutrino  detectors
have been discussed for a long time\cite{john}.
More recent proposals \cite{raghavan} frequently overlook the constraints
imposed by energy-time uncertainty\cite{qm}.  The width of the energy spectrum  reduces to the 
natural line width
only after times of the order of the lifetime of the state emitting the
neutrino. For a shorter time $\delta t$ after the emission of the neutrino, the
observed energy width of the final state energy spectrum is
much brader and given by\cite{qm}

\begin{equation}   
\Gamma_{obs}\approx \frac {\tau}{\delta t}\cdot \Gamma_{nat}
\end{equation} 
where $\Gamma_{obs}$ is the observed width expressed as the second moment of the
energy spectrum, $\tau$ is the lifetime of the decay and $\Gamma_{nat}$ is 
natural line width observed at times longer than the lifetime. The effective
cross section $\sigma_{eff}(\delta t)$ observed at a time $\delta t$ after the
neutrino emission is then 
\begin{equation} 
\sigma_{eff}(\delta t) \approx \frac{\delta t}{\tau}\cdot\sigma_{peak} 
\end{equation} 
where $\sigma_{peak}$ is the
cross section at the peak of the resonance. 

For experiments using long-lived decays like the tritium decay
suggested\cite{raghavan} for use in neutrino detection, $\delta t$ is the time
between the emission of the neutrino and its arrival at the absorber. The
factor $(\delta t/\tau)$ is then sufficiently small to render  this kind of
experiment useless for neutrino detection.                  

The width of the energy spectrum for a single neutrino emitted  in a single
transition evolves in time without interaction with the environment and 
eventually reaches the natural line width. This time-dependent  broadening
differs from other conventional sources of broadening whose widths do not
change with time after emission. These other broadenings generally involve
interactions after emission with the environment, averaging over many slightly
different sharp lines or collective superradiant transitions. None of these
involve a change in the width of the spectrum with time if there are no
subsequent interactions. 

Many  other sources of broadening arise from slight individual differences in
the  conditions for each individual emission. These produce a combination of
lines  each shifted by a slightly different energy to produce an overall
broadened line. This broadened  width does not change with time if there are no
subsequent interactions. 

Broadening that involves an energy-time uncertainty for single events arises in
Dicke superradiance\cite{Super}, where  the spectrum of each single photon
observed has an enhanced width. This differs from the neutrino  case because it
arises from a collective transition which speeds up the lifetime and therefore
increases the natural width. Here the  energy spectrum does not change with
time if there are no interactions. Some examples of modifications of resonance
absorption cross sections for  time-uncertainty broadened incident waves have
been seen in M\"ossbauer spectroscopy with synchrotron  radiation
sources\cite{Gerdau}.

It is a  pleasure to acknowledge discussions with John Schiffer, who called 
my attention to this problem.
%----------------------------------------------------------------------
% This prevents REFERENCES from forcing a page break
%\def\newpage{\vskip10ex}
%
\catcode`\@=11 % This allows us to modify PLAIN macros
\def\references{
\ifpreprintsty \vskip 10ex
%\ifpreprintsty \newpageg
%
\hbox to\hsize{\hss \large \refname \hss }\else
\vskip 24pt \hrule width\hsize \relax \vskip 1.6cm \fi \list
{\@biblabel {\arabic {enumiv}}}
{\labelwidth \WidestRefLabelThusFar \labelsep 4pt \leftmargin \labelwidth
\advance \leftmargin \labelsep \ifdim \baselinestretch pt>1 pt
\parsep 4pt\relax \else \parsep 0pt\relax \fi \itemsep \parsep \usecounter
{enumiv}\let \p@enumiv \@empty \def \theenumiv {\arabic {enumiv}}}
\let \newblock \relax \sloppy
 \clubpenalty 4000\widowpenalty 4000 \sfcode `\.=1000\relax \ifpreprintsty
\else \small \fi}
\catcode`\@=12 % at signs are no longer letters
%-----------------------------------------------------------------
{\tighten

\end{document}